\documentclass[%
reprint,
nofootinbib,
 amsmath,amssymb,
 aps,
pra,
floatfix,
]{revtex4-1}

\usepackage{graphicx}
\usepackage{dcolumn}
\usepackage{bm}
\usepackage{hyperref}
\usepackage{multirow}
\usepackage{amssymb}
\hypersetup{colorlinks=true,
    linkcolor=blue,
    citecolor=blue,
    filecolor=magenta,      
    urlcolor=blue,
    }
\usepackage{chemfig}
\usepackage{soul}
\usepackage[version=4]{mhchem}
\usepackage{subcaption}
\usepackage{graphicx}
\newcommand{\ket}[1]{\left| {#1} \right\rangle}
\usepackage{float}
\NewDocumentCommand{\tens}{e{_^}}{%
  \mathbin{\mathop{\otimes}\displaylimits
    \IfValueT{#1}{_{#1}}
    \IfValueT{#2}{^{#2}}
  }%
}
\DeclareUnicodeCharacter{2212}{-}
\usepackage{pifont}
\begin{document}

\preprint{APS/123-QED}

\title{
Blockwise Optimization for Projective Variational Quantum Dynamics (BLOP-VQD): Algorithm and Implementation for Lattice Systems 
}
\author{Harshdeep Singh}
\email{harshdeeps@kgpian.iitkgp.ac.in}
\affiliation{%
 Center of Computational and Data Sciences, Indian Institute of Technology, Kharagpur, India
}%
\author{Sonjoy Majumder}%
\email{sonjoym@phy.iitkgp.ac.in}
\affiliation{%
  Department of Physics, Indian Institute of Technology, Kharagpur, India}%
\author{Sabyashachi Mishra}%
\email{mishra@chem.iitkgp.ac.in}
\affiliation{%
  Department of Chemistry, Indian Institute of Technology, Kharagpur, India}%

\date{\today}

\begin{abstract}
We present an efficient approach to simulate real-time quantum dynamics using Projected Variational Quantum Dynamics (PVQD), where the computational cost is reduced by strategically optimizing only a subset of the variational parameters at each time step. Typically, the variational ansatz consists of repeated blocks of parameterized quantum circuits, where all parameters are updated in a standard optimization procedure. In contrast, our method selectively optimizes one block at a time while keeping the others fixed, allowing for significant reductions in computational overhead. This semi-global optimization strategy ensures that all qubits are still involved in the evolution, but the optimization is localized to specific blocks, thus avoiding the need to update all parameters simultaneously. We propose different approaches for choosing the next block for optimization, including sequential, random, and fidelity-based updation. We demonstrate the performance of the proposed methods in a series of spin-lattice models with varying sizes and complexity. Our method preserves the accuracy of the time evolution with a much lower computational cost. This new optimization strategy provides a promising path toward high-fidelity simulation of the time evolution of complex quantum systems with reduced computational resources.

\end{abstract}

\maketitle


\section{Introduction}
\label{sec:intro}
\setcounter{equation}{0}
Quantum computing, which harnesses the principles of quantum mechanics, operates in a fundamentally different manner compared to classical computing~\cite{qcintro1, qcintro2}. Unlike classical bits that can only exist in a single state, quantum computing utilizes qubits that can simultaneously occupy multiple states through superposition. Moreover, qubits can become entangled, a phenomenon where the state of one qubit is influenced by the state of another, irrespective of the distance between them. This entanglement, coupled with superposition, enables quantum computers to simultaneously generate and manipulate a vast amount of data, potentially leading to exponential acceleration for specific problems~\cite{Advantage1, farhi2019quantum, jaramillo_2016_quantum, first}. Notably, its ability to catalyze significant transformations is particularly evident in the field of chemistry~\cite{qintro1, qintro3, qintro6, cao_2019_quantum, hd2}. 

Quantum computers are particularly suitable for simulating structure and properties in the areas of catalysis, materials research, and drug discovery~\cite{drug1, drug2, drug3}. In this early era of quantum computing, hybrid classical-quantum algorithms such as Variational Quantum Eigensovlers (VQE) have become one of the most efficient methods for doing molecular electronic structure calculations~\cite{vqa, qintro4, qintro5, hd3, hd4}. These algorithms have undergone experimental testing on several quantum hardware, such as photonic processors (photonic) and trapped-ion processors~\cite{trapion}. The use of efficient ansatzes such as the unitary coupled cluster methods~\cite{qintro2, qintro7}, and the advances in error-resistant algorithms as well as quantum error correction~\cite{error1, error2, error3, halder_2023_development} add further to the utility of these variational algorithms. 
\par
The time evolution of quantum systems is a central challenge in many fields, including condensed matter physics, quantum chemistry, and high-energy physics. The evolution of quantum systems is governed by the time-dependent Schrödinger equation,
\begin{equation}
    i\hbar \frac{d}{dt} |\psi(t)\rangle = H |\psi(t)\rangle.
\end{equation}
The time evolution of a state $|\psi(t)\rangle$, under a time-independent Hamiltonian, can then be given via the time evolution operator $ U(\Delta t) = e^{-iH\Delta t/\hbar}$, as 
\begin{equation}
    U(\Delta t) |\psi(t)\rangle = e^{-i H\Delta t/\hbar} |\psi(t)\rangle = |\psi(t+\Delta t)\rangle.
\end{equation}
Several methods have been devised for studying the time evolution of these quantum systems, including time-dependent density functional theory (TDDFT)~\cite{tddft1, tddft2, tddft3}, Monte-Carlo techniques~\cite{mc1, mc2}, and tensor network methods~\cite{tnqd1, tnqd2}. However, most of these classical methods for simulating quantum dynamics often suffer from the exponential growth of the Hilbert space with system size, making it impossible to simulate large-scale systems efficiently~\cite{qdch1, challenge1, challenge2, challenge3}. 

\par Quantum computing offers a fundamentally different approach to simulating quantum dynamics by inherently utilizing the principles of quantum mechanics, overcoming many limitations of classical methods. The product formula methods, such as Trotter-Suzuki decomposition~\cite{pf1, pf2}, and their modern derivative like qDrift~\cite{qdrift1, qdrift2}, approximate the time evolution of a quantum system by breaking the Hamiltonian into smaller, tractable terms and applying them sequentially. For a Hamiltonian $H = \sum^m_i H_i$, the time evolution operator $e^{-iH \Delta t}$ is approximated using the first-order Suzuki Trotter formula, 
\begin{equation}
    e^{-iH \Delta t} \simeq \bigg( \prod^m_i e^{-i\frac{H_i} {p}\Delta t} \bigg)^p,
\end{equation}
where $p$ is the number of Trotter steps taken for a single time-evolution step $\Delta t$. These smaller steps reduce the overall error introduced by the Trotter approximation. While these methods are efficient and suitable for near-term quantum devices, they introduce errors due to the non-commuting terms $H_i$ in the Hamiltonian. This can be mitigated by higher-order decompositions, albeit with increased circuit depth. 
\par The linear combination of unitaries (LCU) method simulates quantum dynamics by expressing the time evolution operator as a linear combination of unitary operators~\cite{lcu2, lcu1}.
Using techniques like Hamiltonian simulation and oblivious amplitude amplification, LCU prepares the desired state by probabilistically applying these unitaries and post-selecting successful outcomes. Although the LCU method is efficient for sparse Hamiltonians, it requires ancilla qubits and controlled operations, which makes it resource-intensive but powerful for simulating complex quantum systems~\cite{lcumolecule}. 
Quantum walks are a framework for simulating quantum dynamics by modeling the evolution of a quantum state on a graph or a lattice~\cite{quantumwalks}. Quantum Signal Processing (QSP) is a powerful technique for simulating quantum dynamics by applying polynomial transformations to the eigenvalues of a Hamiltonian~\cite{qsp}. It uses a sequence of controlled rotations and phase shifts to approximate functions of the Hamiltonian, like the time-evolution operator. 

\par For near-term devices, however, variational quantum algorithms such as VarQITE (Variational Quantum Imaginary Time Evolution) and variational Hamiltonian simulation~\cite{var_qd_theory, var_qd1, var_qd2, var_qd3}, which classically optimize parameterized circuits to simulate quantum dynamics remain the standard choice. These approaches trade exactness for practicality on noisy hardware. 
Projected variational quantum dynamics (PVQD) is designed to simulate the real-time evolution of quantum systems using parameterized quantum circuits~\cite{pvqd}. This method combines variational principles with global projection techniques to efficiently approximate quantum dynamics. Unlike traditional variational methods, which typically rely on iterative optimization of a restricted subset of parameters, PVQD optimizes all variational parameters simultaneously, enabling it to handle larger and more complex quantum systems.   Other global optimization algorithms, such as those based on the time-dependent variational principle (TDVP)~\cite{var_qd1, var_qd2, var_qd3} suffer from quadratic scaling with the number of parameters, making them unsuitable for large parameterized quantum circuits. On the other hand, PVQD shows an optimal linear scaling with respect to the total number of variational parameters~\cite{pvqd}.

\par In VQAs, the objective function landscape is often highly non-convex and rugged, where the global minimum is not easy to find amid numerous local minima and maxima. The global nature of PVQD, in which all parameters are simultaneously optimized, requires the algorithm to explore a very high-dimensional parameter space to find the optimal set of parameters that minimize the objective function. The optimization landscape becomes even more complex when the number of parameters increases. In such landscapes, gradient-based methods (like those used in classical optimization) can easily get stuck in local minima, making it difficult to identify the optimal set of parameters.

\par This work introduces an efficient approach to updating (sweeping) the variational parameters in the PVQD algorithm by following a divide-and-conquer strategy. Typically, the variational ansatz consists of repeated blocks of parameterized quantum circuits, where all parameters are updated during the variational optimization procedure. Our method selectively optimizes one block at a time while keeping the others fixed, thus significantly reducing computational overhead. This semi-global optimization strategy ensures that all qubits are still involved in the evolution while restricting the optimization to specific blocks, thus avoiding the need to update all parameters simultaneously. We propose different schemes to choose the next block for optimization, including sequential update, random allocation, and fidelity-based choice. These approaches are tested on Ising and Heisenberg lattice models with varying numbers of qubits to demonstrate the computational advantage.



\section{Projected Variational Quantum Dynamics (PVQD)}


To study the evolution of the system under a Hamiltonian $H$ with the PVQD method, the exact quantum state at a given time $t$,
$\vert \phi(t) \rangle$, is approximated to a parameterized ansatz state $\vert \psi_{\theta(t)} \rangle$, where $\theta(t) \in \mathbb{R}^l$ is a vector of parameters $l$. 
The time evolution of the system over a small time-step $\Delta t \in \mathbb{R}$ is governed by the time evolution operator $e^{-i H \Delta t}$.
The state after a small time-step $\Delta t$ is defined as,
\begin{equation}
\vert \phi(t + \Delta t) \rangle = e^{-i H \Delta t} \vert \psi_{\theta(t)}\rangle.
\end{equation}
The system at time $t+\Delta t$ is then mapped to a parameterized state by variationally maximizing the overlap between the evolved state $\vert \phi(t+\Delta t) \rangle$ and the state $\vert \psi_{\theta + d \theta} \rangle$\footnote{Hereafter, $\theta(t)$ is written as $\theta$ to simplify the notation.}, where $d \theta \in \mathbb{R}^l$ is a vector of parameters that represent the change in the state of the system during $\Delta t$ time. Specifically, we want to find $d \theta$ such that
\begin{equation}
    \arg \max_{d \theta \in \mathbb{R}^l} \left| \langle \phi(t+\Delta t) \vert \psi_{\theta+d \theta} \rangle \right|^2. 
\end{equation}
The optimal parameter set ($\theta+\mathbf{d}\theta^\ast$) minimizes the infidelity, defined as
\begin{equation}
    L(\mathbf{d} \theta, \Delta t) = \frac{1 - \left| \langle \phi(t+\Delta t) | \psi_{\theta + \mathbf{d}\theta} \rangle \right|^2}{\Delta t^2}.
\end{equation}
$\Delta t^2$ in the denominator ensures that $L$ becomes independent of the time-step size in the limit $\Delta t \to 0$. In the small time-step limit, the approach is equivalent to McLachlan’s variational principle~\cite{mclachlan}, which ensures that it captures quantum dynamics accurately over time. 
\par
Since the wavefunction is encoded as a quantum circuit $U(\theta)$ acting on some initial state $\ket{0}$, we substitute $\vert \psi_\theta \rangle = U(\theta)\vert 0 \rangle$ into the loss function to obtain
\begin{equation}
    L(\mathbf{d}\theta, \Delta t) = \frac{1 - \left| \langle 0 | U^\dagger(\theta) e^{-i H \Delta t} U(\theta + \mathbf{d}\theta) | 0 \rangle \right|^2}{\Delta t^2},
    \label{loss}
\end{equation}
where the term within modulus represents an expectation value that can be sampled on a quantum computer.
To determine the optimal update $\mathbf{d}\theta^*$, the parameters are iteratively adjusted by moving along the steepest direction of the gradient $\frac{\partial}{\partial \mathbf{d}\theta_i} L(\mathbf{d}\theta, \Delta t)$, starting from an initial guess $\mathbf{d}\theta_0$. 
For a circuit with $l$ variational parameters, each optimization step requires $O(l)$ measurements.  
The optimization continues until the loss function is reduced below the desired threshold.
Once the optimal parameter, updated $\mathbf{d}\theta^*$, is determined, the parameters at time $t + \Delta t$ are updated as:
\begin{equation}
    \theta(t + \Delta t) = \theta(t) + \mathbf{d}\theta^*(t).
\end{equation}

\par
\begin{figure*}[!]
\includegraphics[width=18cm]{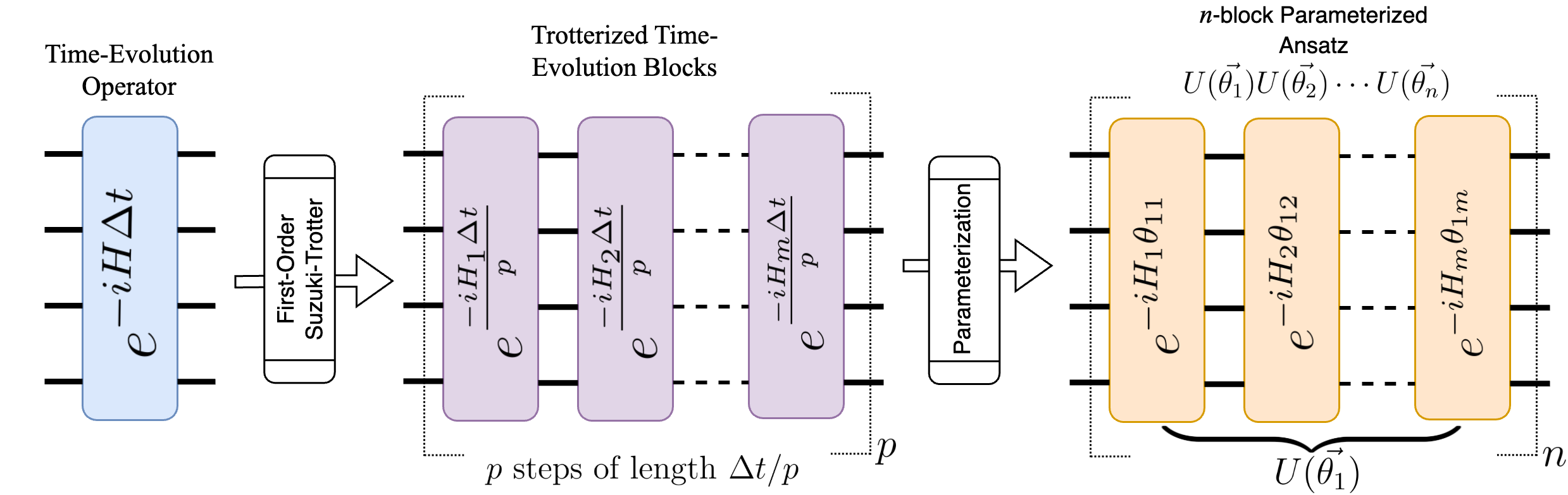}
\caption{\label{fig:blocks}Trotterization of time-evolution operator and ansatz preparation via parameterization of the time-evolution blocks for the PVQD algorithm.} 
\end{figure*}

 The circuit execution of the PVQD algorithm is illustrated in FIG.~\ref{fig:blocks}. The first step involves the construction of a quantum circuit for the time-evolution operator $e^{-iH\Delta t}$ using Trotterization. In this work, we have used the first-order Suzuki-Trotter decomposition with $p$ trotter steps. The ansatz is then constructed by parameterizing the time-evolution blocks and repeating these blocks $n$ times. The values of $p$ and $n$ are problem-dependent (see TABLE~\ref{tab:info} for the systems studied here). 
Notably, in the PVQD method, the circuit’s width and depth remain constant throughout the simulation, and the method does not require auxiliary qubits. However, its global nature, i.e., optimizing all parameters simultaneously, poses challenges, including high-dimensional optimization, sensitivity to initial conditions, and increased susceptibility to noise on NISQ devices.

\subsection{Sweeping Mechanism in PVQD}
For an $N$-qubit system with $m$ gates, the number of variational parameters in the circuit grows as $\mathcal{O}(Nm)$. As $N$ increases, this quickly becomes intractable for classical optimization algorithms that try to forage in the parameter space. As the number of parameters increases, considerable enhancement of computational resources is needed for circuit optimization. In practice, this demands longer runtimes, larger memory utilization, and more frequent (and potentially less accurate) cost function evaluations on quantum devices. This scaling issue can significantly reduce PVQD's effectiveness for large systems or over lengthy simulation periods. Moreover, when optimizing a variational quantum algorithm with many parameters, the cost function may exhibit flat regions or plateaus with small gradients, making progress or convergence difficult. Quantum circuits often do not have simple closed-form gradients. Hence, the optimization algorithms rely on techniques, like parameter shift rules which estimate gradients~\cite{aqgd}. For larger systems with more parameters, these gradient estimations introduce more noise, further exacerbating the difficulty of the optimization process. Even the gradient-free methods (like, genetic algorithms or differential evolution) face challenges due to the large parameter space, apart from being computationally expensive~\cite{Larson_Menickelly_Wild_2019}. 
To overcome the limitations of the standard implementation of the PVQD algorithm, one can construct the parameterized circuit $U(\theta)$ with the help of repetitive blocks (FIG~\ref{fig:blocks}),
\begin{equation}
U^{\otimes n}(\vec{\theta}) = U(\vec{\theta_1})U(\vec{\theta_2})\cdots U(\vec{\theta_n}),    
\end{equation}
where each block $U$ is constructed by parameterizing single-repitition of time-evolution block ($e^{iH\Delta t}$). 

\par
\begin{figure*}[!]
\includegraphics[width=18cm]{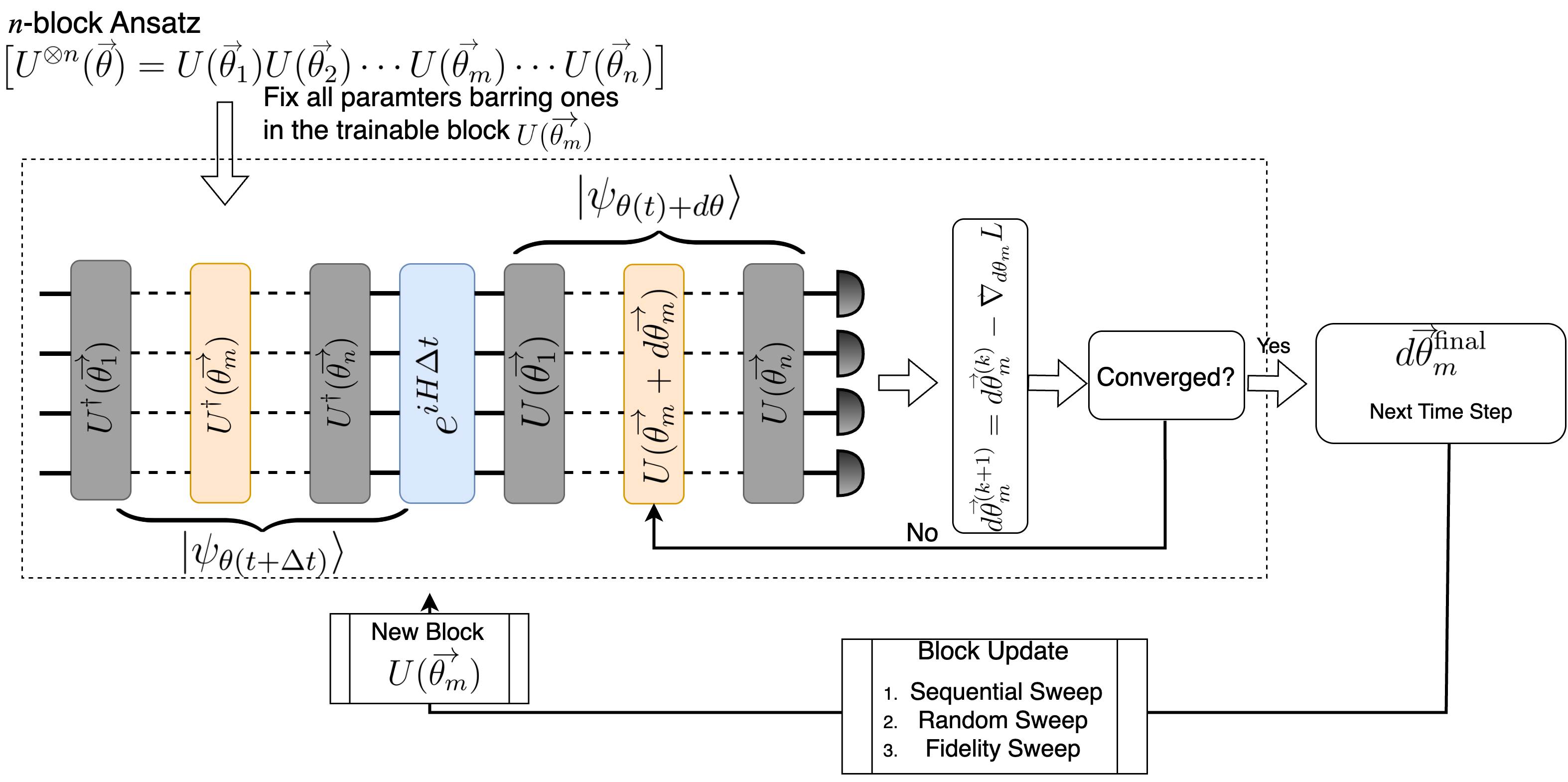}
\caption{\label{fig:workflow}Workflow diagram for PVQD algorithm with efficient sweeping of parameters. }. 
\end{figure*}

 Here we propose that instead of optimizing all the parameters (from all blocks) in the ansatz ($\vec{\theta}$), a more systematic approach can be adopted to update the parameters. 
An efficient sweeping technique allows the parameters of each block among $(\vec{\theta_1}, \vec{\theta_2}, \cdots \vec{\theta_n})$ to be updated one after another, rather than all at once (FIG.~\ref{fig:workflow}). Sequential optimization maintains the global nature of the PVQD algorithm since each block continues to involve the full set of qubits in the system, ensuring that all qubits are entangled and interact with each other at every step of the optimization process. By updating the parameters block by block, it would be possible to reduce the complexity of the optimization task while retaining the capability of the original algorithm to explore the global parameter space effectively. 
Here, we have employed three techniques to choose the different parameterized blocks and sweep the parameters, e.g., \textit{sequential, random}, and \textit{fidelity} sweep (FIG.~\ref{fig:workflow}). In the sequential sweep approach, for every time step, a new block is chosen sequentially for parameter update, i.e.,  $U(\vec{\theta_1})\rightarrow U(\vec{\theta_2})\rightarrow \cdots U(\vec{\theta_n}) \rightarrow U(\vec{\theta_1})$.  In the random sweep approach, for every time step, a new block for parameter update is chosen randomly. On the other hand, in the fidelity sweep method, for every time step, the process starts with updating the parameters from the first block ($U_1$), and the next block is chosen (sequentially) only if the loss function $L$ (Eq.~\ref{loss}) drops below a certain threshold value $L_o$. The fidelity sweep follows a greedy optimization process, where a particular block is kept optimizing until the reward is diminished and then keeps moving to the next available parameterized block.

\subsection{Implementation for Model Systems}
 We explore these protocols with two well-studied spin systems, namely, the transverse field Ising model and the Heisenberg XYZ model, to test the efficiency of different sweeping mechanisms in the PVQD algorithm. The transverse field Ising model represents a quantum spin system that generalizes the classical spin chain by introducing a transverse magnetic field, which causes quantum fluctuations and drives a quantum phase transition between an ordered ferromagnetic phase and a disordered paramagnetic phase. The Hamiltonian for the transverse field Ising model is given by:
\begin{equation}
H = -J \sum_{i} \sigma_i^z \sigma_{i+1}^z - h \sum_{i} \sigma_i^x,   
\end{equation}
where \( \sigma_i^z \) and \( \sigma_i^x \) are the Pauli spin matrices acting on the \(i\)-th spin.
The coupling constant \( J \) determines the strength of the nearest-neighbor interaction along the \( z \)-axis, while \( h \) is the strength of the transverse magnetic field acting on each spin along the \( x \)-axis. Here, we have chosen a ferromagnetic system with $J = -0.25$ and $h = -1$. In Ising model, the observables studied are the magnetization along $X$, and $Z$ axis, defined by,
\begin{equation}
    \sigma_x = \sum^{N}_i X_i, \quad \sigma_z = \sum^{N}_i Z_i, 
\end{equation}
where, $X_i$, and $Z_i$ are the Pauli operations on the $i^{th}$ qubit in the $N$ qubit system.
 
The Heisenberg XYZ model is a quantum spin model that generalizes the classical Heisenberg model by introducing anisotropic interactions between neighboring spins in all three spatial directions. The Hamiltonian for the model is given by 
\begin{equation}
H = \sum_{i} \left( J_x \sigma_i^x \sigma_{i+1}^x + J_y \sigma_i^y \sigma_{i+1}^y + J_z \sigma_i^z \sigma_{i+1}^z \right)    
\end{equation}
where \( \sigma_i^x \), \( \sigma_i^y \), and \( \sigma_i^z \) are the Pauli matrices acting on the \(i\)-th spin, and \( J_x \), \( J_y \), and \( J_z \) are the coupling constants for the interaction between neighboring spins along the \( x \), \( y \), and \( z \)-axes, respectively. In this work, we have chosen $J_x = 1.0, J_y = 0.8, J_z = 0.6$, representing a long-wave anisotropic XYZ-model. In the Heisenberg model, the observables studied are $Z_0$, and $Z_0Z_1$, which are the Pauli $Z$ operations at the specified sites. The $Z_0$ represents the local magnetization or spin projection along the $z-$axis at site $0$, while $Z_0Z_1$ represents the two-point correlation between spins at sites $0$ and $1$ along the $z-$axis.

\par Based on the results of our previous work~\cite{hd}, L\_BFGS\_B was chosen as the classical optimizer in the case of ideal quantum settings, while SPSA was chosen as the optimizer for the noisy quantum settings. The numbers of qubits, Trotter steps, ansatz repetitions, and variational parameters for different systems are presented in TABLE~\ref{tab:info}. 

            

\begin{table}[h]
    \centering
    \begin{tabular}{|c|c|c|}
        \hline
        Method  & \# Ansatz Blocks ($n$)  & \# Parameters  \\
        \hline    
        \multicolumn{3}{|c|}{Ising Model with 8 qubits, 15 terms, 8 steps}  \\ \hline
             PVQD(1)  & 1 & 23 \\
             PVQD(2)  & 2 & 46 \\
             FS(2)       & 2 & 46 (23) \\
        \hline
        \multicolumn{3}{|c|}{Ising Model with 10 qubits, 19 terms, 8 steps}  \\ \hline
            PVQD(1)  & 1 & 29 \\
             PVQD(2)  & 2 & 58 \\
             PVQD(4)  & 4 & 116 \\
             FS(2)       & 2 & 58(29)\\
        \hline 
        \multicolumn{3}{|c|}{Heisenberg Model with 10 qubits, 27 terms, 8 steps}  \\ \hline
            PVQD(1)  & 1 & 54 \\
             PVQD(2)  & 2 & 108 \\
             PVQD(4)  & 4 & 216 \\
             FS(2)    & 2 & 108 (54) \\
             FS(4)    & 4 & 216 (54) \\
        \hline
    \end{tabular}
    \caption{Numbers of qubits, Trotter steps, ansatz repetitions, and the total number of variational parameters (with the number of parameters being optimized given in the parentheses) for different spin systems used in our calculations. }
    \label{tab:info}
\end{table}

\section{Results and Discussion}

\subsection{Performance under Ideal Condition}
\begin{figure*}[!]
\includegraphics[width=18cm]{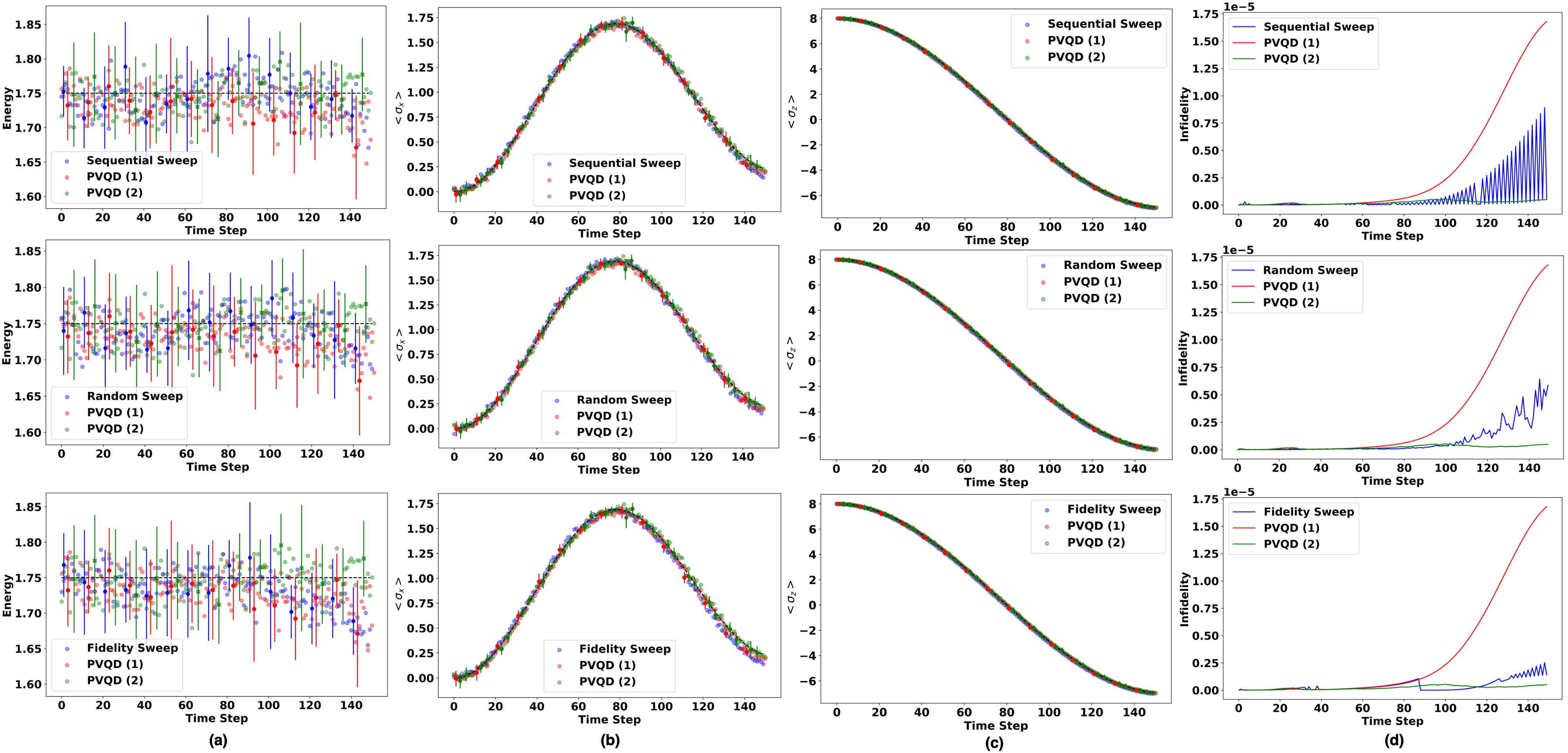}
\caption{\label{fig:ising8q}Average values and standard deviations of (a) Energy, (b) $\langle \sigma_x \rangle $, (c) $\langle \sigma_z \rangle $, (d) Infidelity across 10 simulation runs for a 8-qubit Ising model with different sweeping methods against the standard PVQD. The dotted line in each figure highlights the exact values of the observables. Each time step is 0.01 units.}
\end{figure*}

The dynamics of spin systems can be analyzed from the variations of magnetization parameters over time under noiseless conditions.
FIG.~\ref{fig:ising8q} shows the average values and standard deviations of energy, magnetization along $x-$axis ($\langle \sigma_x \rangle $), magnetization along $z$-axis ($\langle \sigma_z \rangle $), and infidelity over 10 simulation runs under ideal quantum settings for an 8-qubit Ising model. The conventional PVQD runs with one (PVQD(1)) and two repetition blocks (PVQD(2)) are compared with all three different sweeping approaches (i.e., sequential, random, and fidelity) in FIG.~\ref{fig:ising8q}(d). The PVQD(2) method, with two repetition blocks, has a deeper circuit depth and twice the parameters to optimize compared to PVQD(1). With greater computational cost, PVQD(2) achieves high fidelity that remains unmatched by PVQD(1). Unless mentioned otherwise, all the sweeping methods employed in this work have two repetition blocks. Therefore, they have the circuit depth equivalent to PVQD(2), while the number of parameters to optimize is equivalent to PVQD(1).

FIG.~\ref{fig:ising8q}(a, b, c) show the excellent agreement of energy and magnetizations of the different sweeping methods with standard PVQD. The fluctuations in the energy measurement (FIG.~\ref{fig:ising8q}(a)) are due to the projective nature of the PVQD algorithm. All three sweeping methods outperform PVQD(1), see FIG.~\ref{fig:ising8q}(d). The sequential and random sweep approaches do not retain high fidelity and show large fluctuations on longer timescales. However, the fidelity sweep method continues to provide high fidelity comparable to that in PVQD(2), even at longer timescales (FIG.~\ref{fig:ising8q}(d)). Hence, only the fidelity sweep method was employed for the larger systems, e.g., the Ising model with 10 qubits and the Heisenberg model with 10 qubits.

\begin{figure}[!]
\includegraphics[width=0.5\textwidth]{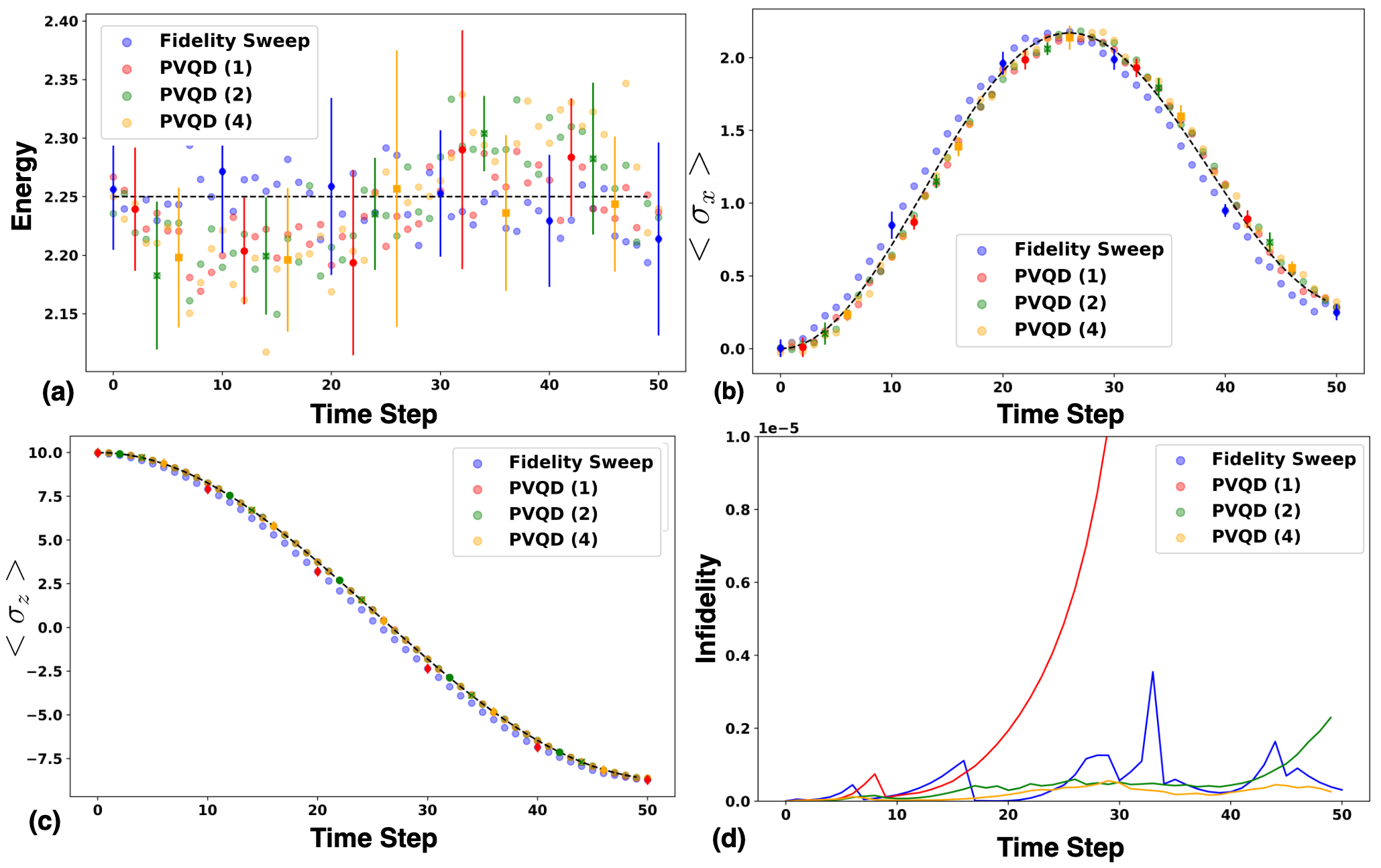}
\caption{\label{fig:ising10q}Average values and standard deviations of (a) Energy, (b) $\langle \sigma_x \rangle $, (c) $\langle \sigma_z \rangle $, (d) Infidelity across 10 simulation runs for a 10-qubit Ising model with fidelity sweep technique against the standard PVQD. The dotted line in each figure highlights the exact values of the observables. Each time step is 0.03 units.}
\end{figure}

\begin{figure}[!]
\includegraphics[width=0.5\textwidth]{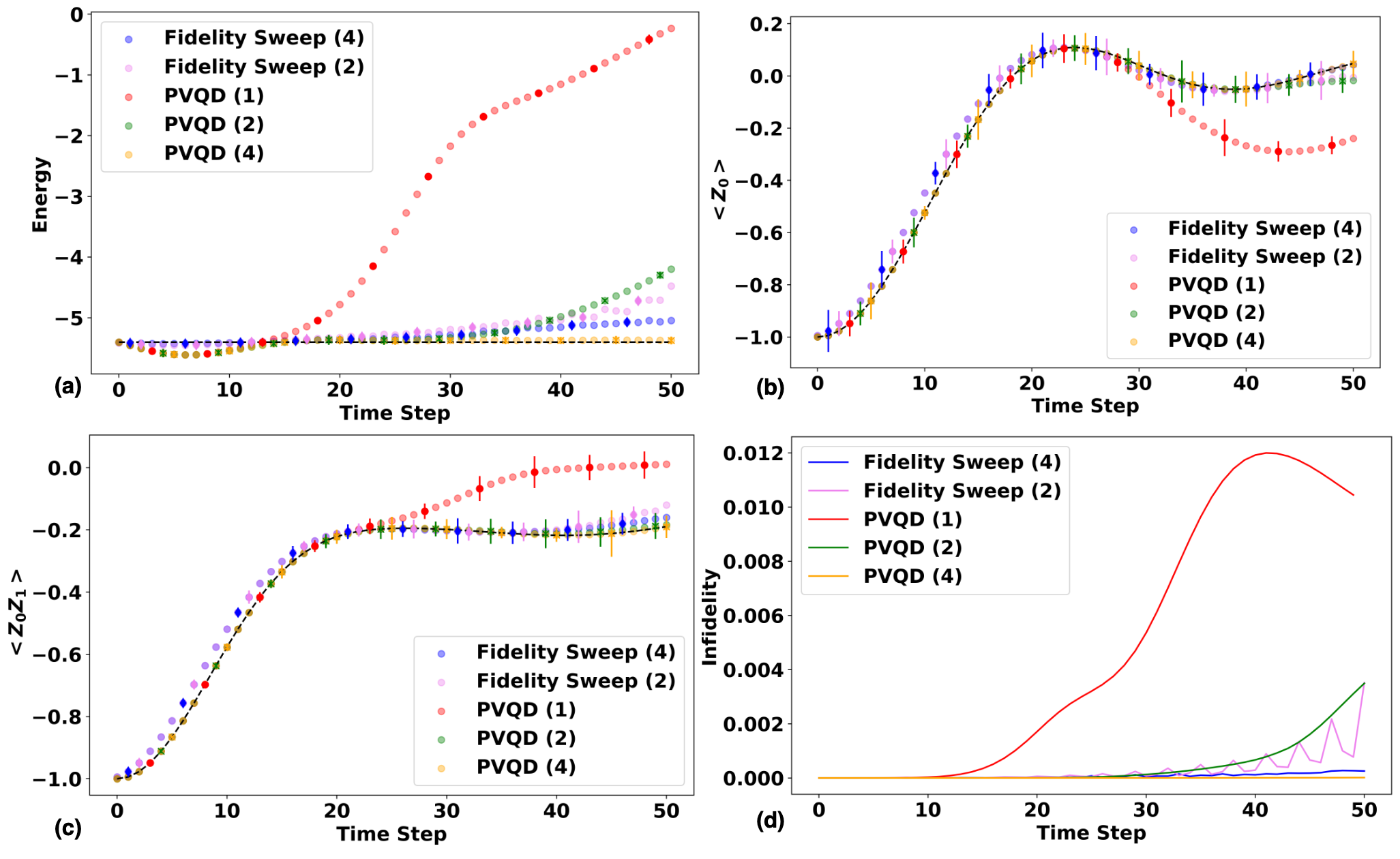}
\caption{\label{fig:heisen10q}Average values and standard deviations of (a) Energy, (b) $\langle Z_0 \rangle $, (c) $\langle Z_0Z_1 \rangle $, (d) Infidelity across 10 simulation runs for a 10-qubit Heisenberg-XYZ model with fidelity sweep technique against the standard PVQD. The dotted line in each figure highlights the exact values of the observables. Each time step is 0.03 units.}
\end{figure}

FIG.~\ref{fig:ising10q} shows the average values and standard deviations of different observables for a 10-qubit Ising system. In this case, PVQD (1) and PVQD (2) perform poorly on longer timescales. It took four repetition blocks (i.e., PVQD(4)) to retain high fidelity at longer timescales. Meanwhile, the fidelity sweep method with two repetition blocks achieves an accuracy comparable to PVQD(4), employing only a quarter of the parameters present in the latter. This improved performance can be further observed for more complicated systems such as the Heisenberg-XYZ spin systems (FIG.~\ref{fig:heisen10q} d). In this case, standard PVQD(1) struggles with large errors for all observables. With the same number of training parameters, fidelity sweep with two repetition blocks (Fidelity Sweep (2)) and four repetition blocks (Fidelity Sweep (4)) provides a much superior performance, massively outperforming PVQD(1) with same number of training parameters and similar performance to PVQD(2) and PVQD(4). The exact values of the errors in energy and different observables for all these systems are given in TABLE~\ref{tab:error_ideal}, showing that the fidelity sweep can evaluate energy and different observables with similar accuracy to PVQD(2) and PVQD(4) methods, and improve the fidelity results by a couple of orders against the PVQD(1) scheme.

\begin{table*}[htbp]
\caption{\label{tab:error_ideal} The average error ($\sum^{N_t}_i|\langle O^i_{\text{exact}}-O^i_{\text{simulated}}\rangle|$) and standard deviation of energy and different observables for different systems considered in the study under ideal quantum settings.}
\begin{ruledtabular}
\begin{tabular}{c|c|c|c|c|c}

System   & Method  & $\Delta \langle E \rangle $  & $\Delta \langle \sigma_x \rangle $/$\Delta \langle Z_0 \rangle $  & $\Delta \langle \sigma_z \rangle $/$\Delta \langle Z_0Z_1 \rangle $  & $\Delta \mathcal{F} $   \\
\hline



Ising (8)  & Fidelity Sweep & $0.02402 \pm 0.05749 $      & $0.03767 \pm 0.05922$  &   $0.01172 \pm 0.04375$  & $4.13\times 10^{-7} \pm 9.84\times 10^{-10}$\\

         & PVQD(1)  &  $0.02434 \pm 0.05571 $  & $0.02284 \pm 0.06034 $  & $0.01369 \pm 0.04346 $ & $3.34\times 10^{-6} \pm 3.37\times 10^{-8}$\\
         & PVQD(2)  &  $0.01924 \pm 0.05697 $  & $0.02238 \pm 0.05738 $  & $0.01489 \pm 0.04361 $ & $2.32\times 10^{-7} \pm 1.35\times 10^{-8}$\\

\hline
Ising (10)  & Fidelity Sweep(4) & $0.01932 \pm 0.06408 $      & $0.04926 \pm 0.06984$  &   $0.03699 \pm 0.0492$  & $5.19\times 10^{-7} \pm 8.92\times 10^{-11}$\\
         & PVQD(1)  & $0.05821 \pm 0.06651 $      & $0.05996 \pm 0.06294$  &   $0.03461 \pm 0.04873$  & $3.65\times 10^{-5} \pm 8.52\times 10^{-8}$\\

         & PVQD(2)  &  $0.03968 \pm 0.06304 $  & $0.04098 \pm 0.06432 $  & $0.01383 \pm 0.04930 $ & $4.63\times 10^{-7} \pm 6.99\times 10^{-9}$\\

         & PVQD(4)  &  $0.01733 \pm 0.02134 $  & $0.04001 \pm 0.06744 $  & $0.01124 \pm 0.03771 $ & $2.16\times 10^{-7} \pm 7.02\times 10^{-9}$\\
\hline

Heisenberg (10)  & Fidelity Sweep(2) & $0.20142 \pm 0.05078 $      & $0.02856 \pm 0.05106$  &   $0.02510 \pm 0.03777$  & $0.00024 \pm 8.88\times 10^{-6}$\\
       & Fidelity Sweep(4) & $0.11809 \pm 0.05492 $      & $0.02619 \pm 0.05751$  &   $0.02204 \pm 0.04403$  & $6.98\times 10^{-5} \pm 8.28\times 10^{-9}$\\

         & PVQD(1)  &  $2.09840 \pm 0.05085 $  & $0.07595 \pm 0.05058 $  & $0.07792 \pm 0.03815 $ & $0.00479 \pm 1.88\times 10^{-8}$\\

         & PVQD(2)  &  $0.23273 \pm 0.04995 $  & $0.03541 \pm 0.03999 $  & $0.02646 \pm 0.03829 $ & $0.00040 \pm 1.13\times 10^{-7}$\\
         & PVQD(4)  &  $0.05734 \pm 0.05697 $  & $0.01745 \pm 0.05428 $  & $0.01003 \pm 0.04361 $ & $3.11\times 10^{-6} \pm 1.35\times 10^{-7}$\\

\end{tabular}
\end{ruledtabular}
\end{table*}

\subsubsection{Computational Cost of different methods}
The number of parameters that need to be trained for a standard PVQD algorithm depends on several factors. The building of the ansatz requires variational parameters in the time-evolution blocks. These time-evolution blocks are typically obtained from Trotterization (to the order $p$) of the problem Hamiltonian $H$ that may contain $m$ terms. In such a case, the number of blocks in the Trotter expansion is $p m$. The number of terms in the Hamiltonian is problem-specific. For a $N$-qubit molecular system, it can scale as $\mathcal{O} (N^4)$, while it can scale as $\mathcal{O} (N^2 \sim N^3)$ in a lattice system depending on the interaction and dimension of the lattice. In an $n$-times repeated variational ansatz, the number of variational parameters scales as $nm$.  If the total number of steps in the time evolution is $N_t$, the number of variational iterations per time step is $N_i$, the number of measurements (or shots) per time step is $N_s$, and the time complexity for each optimization step is $T$, then the overall time complexity of the problem becomes $\mathcal{O} (n m N_t N_i N_s T)$. In the PVQD method, the number of iterations for each time step scales linearly with the number of parameters in the circuit (i.e., $N_i \sim \mathcal{O}(l)$)~\cite{pvqd}. The classical optimizer that involves computing the Hessian or approximations of the Hessian (e.g., L\_BFGS\_B used in our case) can have a time complexity of $l^2$. This makes the overall time complexity of the algorithm $\mathcal{O}(n^{4} m^4 N_t N_s)$. 
\par
\begin{figure}[!]
\includegraphics[width=0.5\textwidth]{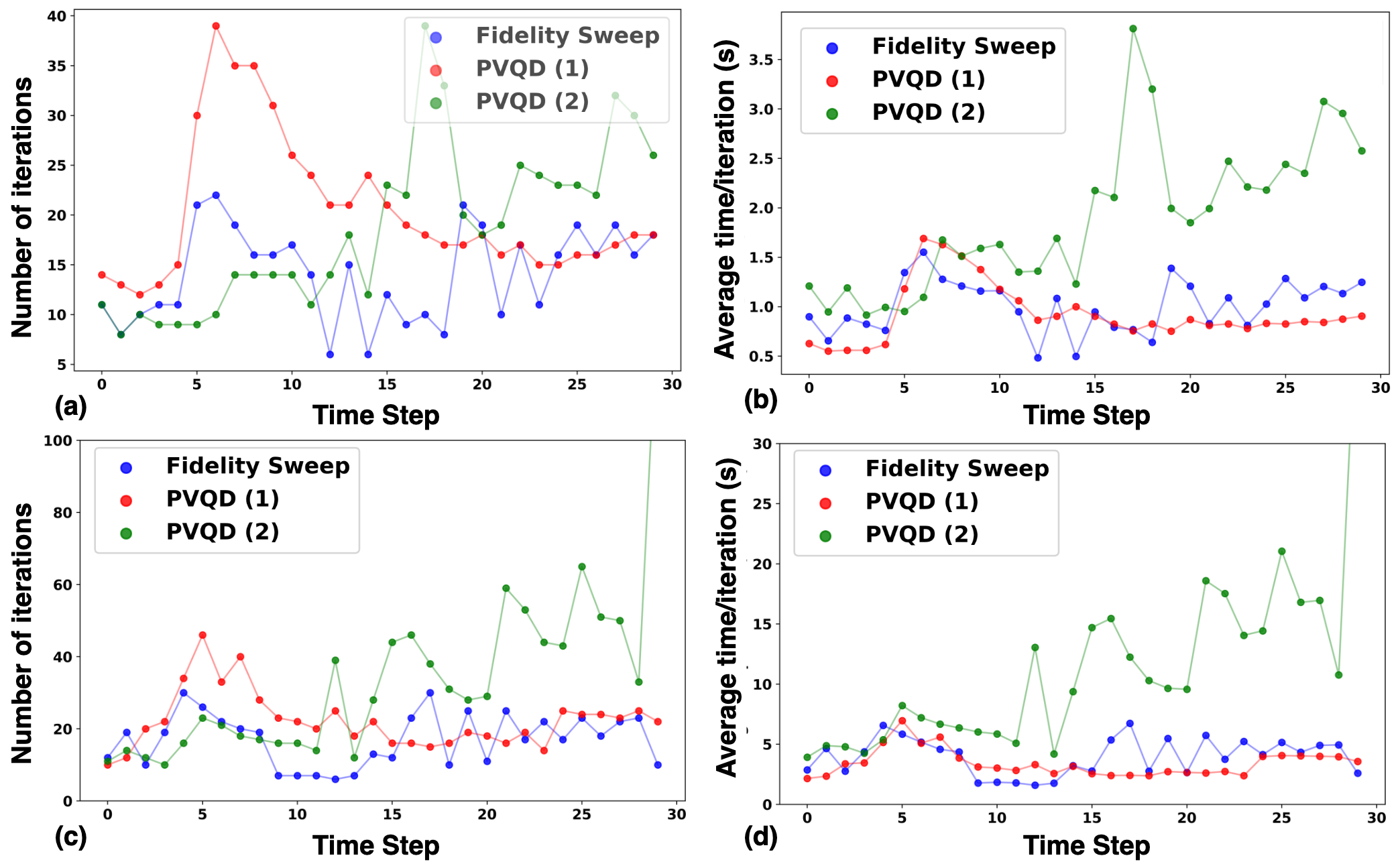}
\caption{\label{fig:isingtime}The average number of iterations and the average time taken per iteration (in seconds) for a 4- qubit [(a), (b)] and 8-qubit Ising systems [(c), (d)] over 10 simulation runs. The time step is 0.05 units.}
\end{figure}

The computational cost decreases considerably when PVQD is performed using the efficient sweeping methods described in this work. The time complexity of the algorithm drops to $\mathcal{O} (m^4 N_t N_s)$ because only one block is explicitly optimized ($n=1$) at any time. Now, generally, the number of blocks in the variational ansatz, if constructed by parameterizing the time-evolution block, is directly dependent on the number of terms in the Hamiltonian. Hence, in those cases, $n\sim\mathcal{O}(m)$, and therefore, by employing the sweeping method, the overall complexity of PVQD can be reduced from $\mathcal{O} (m^8 N_t N_s)$ to $\mathcal{O} (m^4 N_t N_s)$. 
This is further highlighted in FIG.~\ref{fig:isingtime}, which shows the average number of iterations and the average time taken per iteration for the 4- and 8-qubit Ising systems. 
As expected, the number of iterations and the time-per-iteration increase with the number of qubits and parameters in the system. Interestingly, however, the average time taken within a single simulation run in PVQD(2) also increases with time, with the latter half of the simulation taking longer than the first.
Both problems are efficiently tackled using the PVQD method with the fidelity sweep algorithm. The extent of the efficiency of the fidelity can further be highlighted by comparing the run time for the 10-qubit Heisenberg system. The run-time for a single simulation increases from 56 minutes with PVQD(1) to 98 and 1322 minutes for PVQD(2) and PVQD(4), respectively. Whereas Fidelity Sweep (2) and Fidelity Sweep (4) take 61 and 117 minutes, respectively\footnote{All the simulation times are on an Apple Silicon M1 chip device}. This illustrates the efficiency of our PVQD algorithm, which saves run-time and memory by choosing the sweeping method.

\subsubsection{Modification for longer time evolutions} 
\begin{figure}[!]
\includegraphics[width=0.5\textwidth]{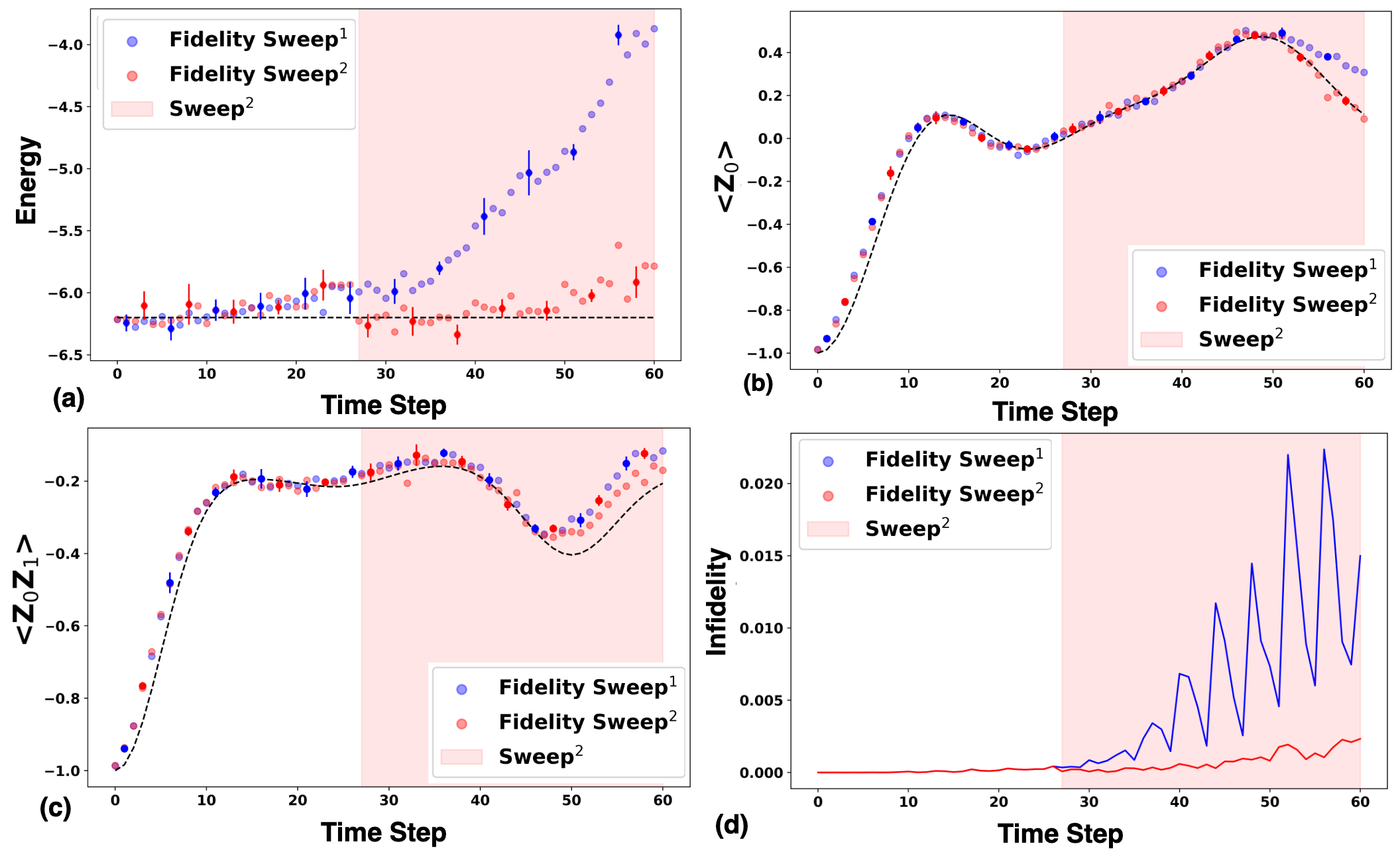}
\caption{\label{fig:2sweep}Average values and standard deviations of (a) Energy, (b) $\langle Z_0 \rangle $, (c) $\langle Z_0Z_1 \rangle $, (d) Infidelity across 10 simulation runs for a 12-qubit Heisenberg-XYZ model with fidelity sweep$^1$, and sweep$^2$ techniques. The shaded red region highlights the intervals where two variational blocks were simultaneously updated.  The dotted line in each figure highlights the exact values of the observables. Each time step is 0.05 units.}
\end{figure}

Rapid fluctuations in the infidelity are found to appear for larger systems at a longer timescale of the simulation (see Fig.~\ref{fig:ising8q} and Fig.~\ref{fig:heisen10q}). This indicates that optimizing a single variational block might not be enough to meet demands of high accuracy in the results during the longer evolutions of large systems. A simple modification can be made to the sweeping pattern where one can slowly increase the number of variational blocks updated at a time, denoted by Sweep$^n$, where $n$ is the number of blocks being trained simultaneously. The blocks can again be chosen sequentially or randomly based on the fidelity calculations. This is demonstrated in FIG.~\ref{fig:2sweep}, which highlights the average values and standard deviations of the different observables across 10 simulation runs of 80 time-steps ($\Delta t =0.05$) each for a 12-qubit Heisenberg-XYZ model with fidelity sweep$^1$, and sweep$^2$ techniques. For the initial part of the time evolution, only one variational block was updated, and once no further improvement could be realized with sweep$^1$, sweep$^2$ is activated. The shaded red region highlights the intervals, where two variational blocks are updated simultaneously.

\subsection{Performance under Noisy Condition}
\begin{figure*}[!]
\includegraphics[width=18cm]{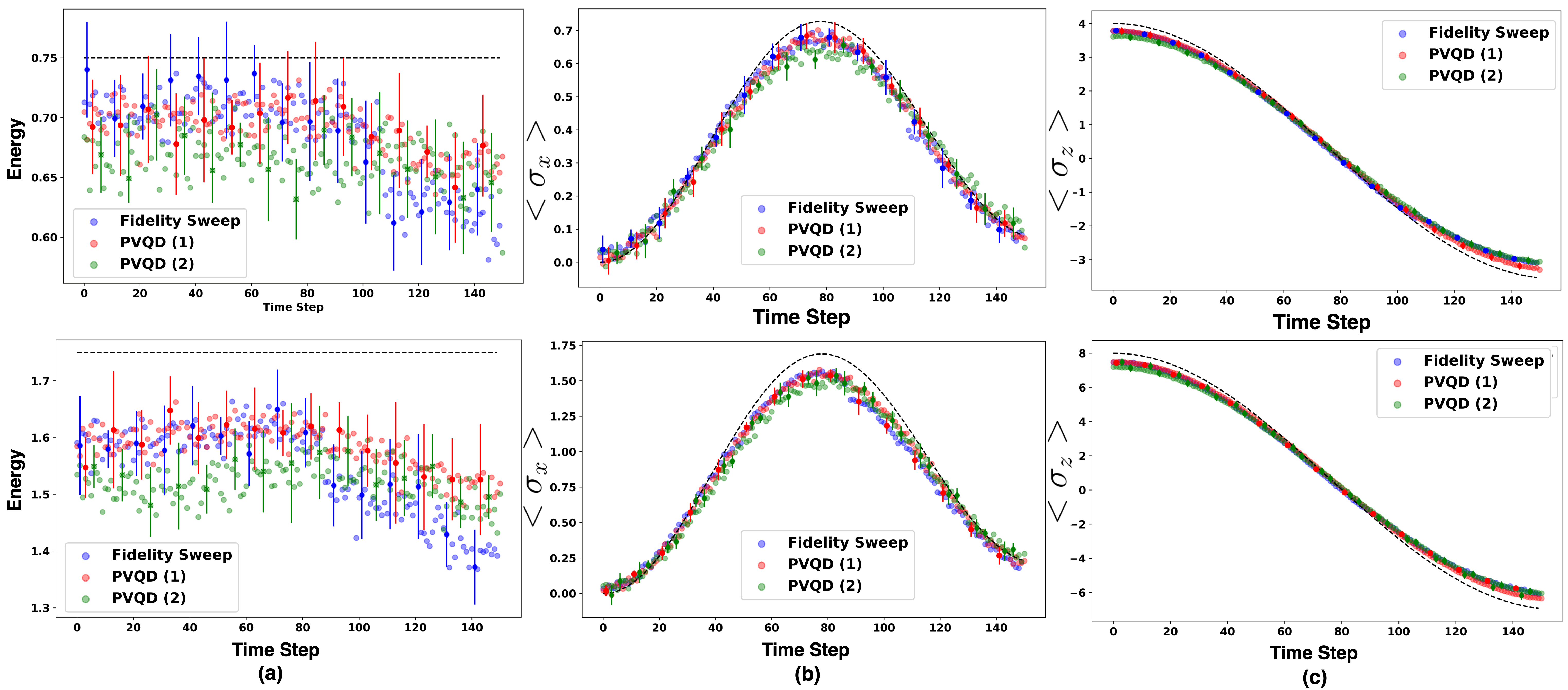}
\caption{\label{fig:isingnoisy}Average values and standard deviations of (a) Energy, (b) $\langle \sigma_x \rangle $, and (c) $\langle \sigma_z \rangle $ across 10 simulation runs under noisy quantum settings for a 10-qubit Ising model with fidelity sweep technique against the standard PVQD. The dotted line in each figure highlights the exact values of the observables. Each time step is 0.03 units.}
\end{figure*}

The physical quantum computer is prone to noise and it is necessary to develop a better algorithm abiding with it.  FIG.~\ref{fig:isingnoisy} shows the average values and standard deviations of energy, magnetization along $x-$axis ($\langle \sigma_x \rangle $), and $z-$axis ($\langle \sigma_z \rangle $) across 10 simulation runs under realistic noisy quantum settings for a 4-qubit and 8-qubit Ising model with fidelity sweeping method against the standard PVQD with one (PVQD(1)) and two (PVQD(2)) repetition blocks. Just like in the previous works~\cite{pvqd, adaptpvqd}, the variational parameters are evaluated on an ideal simulator and the expectation values of the observables are evaluated in the noisy settings. As expected, in the presence of noise, the error in the expectation values increases significantly for all the different methods. This is further highlighted in TABLE~\ref{tab:error}. Interestingly, the highest error is reported in the PVQD(2) method since it has the highest number of training parameters and the circuit depth. However, in the fidelity sweep method, the number of training parameters reduction results in a smaller error across different observables, even though it has the same circuit depth as PVQD(2). Hence, efficient sweeping not only helps in saving computational resources but also increases resilience against the quantum noise of the PVQD algorithm.

\begin{table}[htbp]
\caption{\label{tab:error} The average error ($\sum^{N_t}_i|\langle O^i_{\text{exact}}-O^i_{\text{simulated}}\rangle|$) and standard deviation of energy and magnetization for different Ising systems under noisy quantum settings.}
\begin{center}
\begin{ruledtabular}
\begin{tabular}{c|cccc}

System  & Method  & $\Delta \langle E \rangle $  & $\Delta \langle \sigma_x \rangle $  & $\Delta \langle \sigma_z \rangle $   \\
\hline
Ising $(4q)$ & Fidelity Sweep & $0.06663$      & $0.03068$  &   $0.20397$\\
             &                & $\pm 0.04054 $  & $\pm 0.04169 $  &  $\pm 0.03328 $   \\
      & PVQD(1)  &  $0.05956$  & $0.02694$  & $0.15583$\\
      &        & $\pm 0.04278 $ & $\pm 0.04193 $  &  $\pm 0.03268 $   \\
     & PVQD(2)  &  $0.08469$  &  $0.041692$ & $0.25989$\\
    &          & $\pm 0.04145 $    &  $\pm 0.04057$   &  $\pm 0.03408 $   \\
\hline
Ising $(8q)$ & Fidelity Sweep & $0.20758$      & $0.09228$  &   $0.46220$\\
             &                & $\pm 0.06881 $  & $\pm 0.06113 $  &  $\pm 0.09950 $   \\
      & PVQD(1)  &  $0.16537$  & $0.07028$  & $0.34823$\\
      &        & $\pm 0.06831 $ & $\pm 0.05989 $  &  $\pm 0.09323 $   \\
     & PVQD(2)  &  $0.23002$  &  $0.09897$ & $0.53638$\\
    &          & $\pm 0.06206 $    &  $\pm 0.05511$   &  $\pm 0.09408 $   \\

\end{tabular}
\end{ruledtabular}
\end{center}
\end{table}

\subsection{Warm Starting the Optimization Blocks}
\begin{figure*}[!]
\includegraphics[width=18cm]{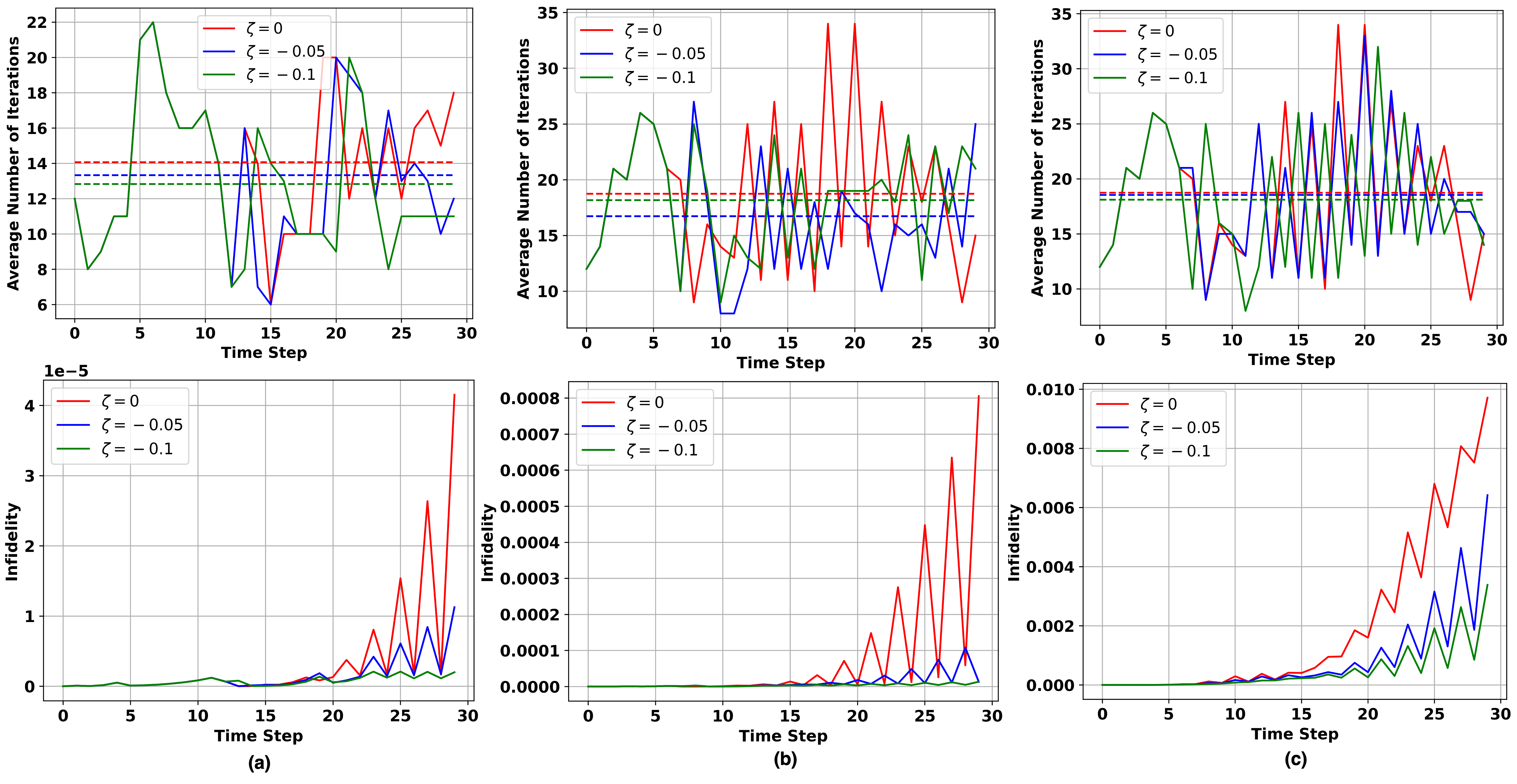}
\caption{\label{fig:warmstart}The average number of iterations and the infidelity for (a) 4- qubit (b) 8-qubit Ising systems, and (c) 8-qubit Heisenberg system with different values of $\zeta$. The time step is 0.05 units.}
\end{figure*}

So far in this work, all initial parameters of the different variational blocks were set to zero for all simulations. 
Consider the case where a certain variation block is updated at a given time step and a different block needs to be updated in the next step.  In the current implementation, the optimized parameters of the previous block do not influence the initial parameters of the new block that need to be updated. However, one can impart some information from the optimized parameters of the previous block to the new block as a method of warm-starting (i.e., efficient initialization) the parameters of the next block. Suppose that the current optimization block $U_i (\vec{\theta_i})$ is updated with the parameters $\vec{\theta_i^*}$. In the next time step, if a different block ($U_j (\vec{\theta_j})$) is chosen for optimization, then the parameters of block $U_j$ are initialized as,
\begin{equation}
    \vec{\theta_j} = \vec{\theta_j} + \zeta \vec{\theta^*_i}
\end{equation}
where $\zeta$ acts as a secondary learning rate. FIG.~\ref{fig:warmstart} highlights the average number of iterations and the infidelity for 4-qubit and 8-qubit Ising systems over 10 simulation runs. It shows that for $\zeta=-0.05$ and $-0.1$, the average number of iterations can be reduced by 10\% as compared to the standard initialization (i.e., all parameters set to zero). A random initialization of the parameters leads to much worse results. Although different values of $\zeta$ were considered (data not shown), $\zeta=-0.05$ and $-0.1$ produced the best results. Further, the warm starting of the next optimization blocks also leads to much better infidelity results, especially in the later parts of the time evolution (FIG.~\ref{fig:warmstart}).

\section{Conclusions}
This work introduces a novel approach to simulating real-time quantum dynamics using Projected Variational Quantum Dynamics (PVQD) with a semi-global block optimization strategy. By selectively optimizing one block of the variational ansatz at a time while keeping others fixed, we significantly reduced the computational overhead without compromising the accuracy of the time evolution. Our method ensures that all qubits remain actively involved in the dynamics while localizing the optimization process, thereby avoiding the resource-intensive task of updating all parameters simultaneously. 
The dynamics of various lattice systems with varying numbers of qubits demonstrate that using an efficient sweeping algorithm can save a significant amount of run-time and memory. The fidelity-based approach to sweeping the parameterized blocks is found to be more efficient than sequential or random updates. The efficient sweep method reduces the time complexity of a $N-$qubit Hamiltonian with $m$ terms from $\mathcal{O} (m^8 N_t N_s)$ to $\mathcal{O} (m^4 N_t N_s)$. The warm-starting choice of the parameters is seen to improve the algorithm's performance further. The results highlight the potential of this method to enable high-fidelity quantum dynamics simulations with reduced computational resources, paving the way for practical applications in quantum chemistry, materials science, and beyond. This work underscores the importance of innovative optimization strategies in harnessing the power of variational quantum algorithms for realistic quantum simulations.

\section*{Conflict of interest}
There are no conflicts to declare.

\section*{Acknowledgements}
This work used the Supercomputing facility of IIT Kharagpur established under the National Supercomputing Mission (NSM), Government of India, and supported by the Centre for Development of Advanced Computing (CDAC), Pune. HS acknowledges the Ministry of Education, Govt. of India, for the Prime Minister's Research Fellowship (PMRF).

\section*{Author Contribution} 
HS: Data Curation, Formal Analysis,  and  Original Draft Writing. S. Majumder and S. Mishra: Supervision, Review \& Editing.

\bibliography{main.bib}

\end{document}